\documentclass{amsart}
\newtheorem{theorem}{Theorem}[section]

\theoremstyle{definition}

\theoremstyle{remark}

\numberwithin{equation}{section}

\renewcommand{\Re}{\mathbb R}          

\newcommand{\tr}{{\text{\rm tr}}}
\renewcommand{\d}{\partial}

\newcommand{\half}{\frac{1}{2}}
\newcommand{\ordo}{o}
\newcommand{\ame}{{}^{(4)}g}            
\newcommand{\aR}{{}^{(4)}R}             
\newcommand{\ok}{{}^0k}                 
\newcommand{\og}{{}^0g}                 
\newcommand{\ophi}{{}^0\phi}            
\newcommand{\SR}{{}^S R}                

\title{Quiescent cosmological singularities}
\author{Lars Andersson}
\address{Department of Mathematics\\
Royal Institute of Technology\\
100 44 Stock\-holm, Sweden\\
}
\email{larsa@math.kth.se}
\thanks{Talk given at ICMP2000, London. 
Supported in part by the Swedish Natural
Sciences Research Council (SNSRC),  contract no.  R-RA 4873-307}

\date{January 1, 1994 and, in revised form, June 22, 1994.}

\begin{document}

\maketitle

\section{Introduction}\label{sec:intro}
This is a report on joint work with Alan Rendall \cite{andersson:rendall}, 
on non--oscillatory singularities in four dimensional space--times with
scalar field or stiff fluid matter. Before considering the results of 
\cite{andersson:rendall} in detail, I will discuss the 
BKL proposal which motivated the work. 

The singularity theorems of Penrose and Hawking guarantee the existence 
of spacetime singularities under reasonable assumptions,  but give little
information about the singularities they predict. 
The most detailed proposal for the structure of space--time singularities is
due to Belinskii, Khalatnikov and Lifshitz 
(BKL), see \cite{lifshitz63,bkl70,bkl82} and references therein. According to 
the BKL proposal, singularities in generic four dimensional 
space--times are space--like and
oscillatory, while generic 
space--times with stiff fluid, 
including
massless scalar fields, have singularities which,
according to a proposal of Belinskii and Khalatnikov, see
\cite{belinskii73}, are space--like and non--oscillatory.
Further, generic space--times will, according to this picture, have the
property that spatial points decouple near the singularity, and the local
behavior is 
asymptotically like spatially homogeneous (Bianchi) models locally near the
singularity. In particular, in the non--oscillatory case, spatial
derivatives become unimportant, the term ``asymptotically velocity term
dominated'' is often used to describe this phenomenon. Space--times with  
non--stiff matter appear to have 
the property that asymptotically
close to the singularity, matter becomes insignificant and the main features
of the evolution is determined by curvature. On the other hand, in the case
of stiff matter, the picture is that the matter becomes dominant near the
singularity, leading to non--oscillatory behavior. We will refer to the
collection of ideas described above as the BKL proposal. 

The Bianchi models are space--times with spatially locally homogeneous
geometry. This is an interesting toy model for the structure of
singularities, and the so--called mixmaster universe, Bianchi IX, has been
used 
as a model for the typical local behavior, near the
singularity, of gravity in four dimensions, with non-stiff
matter. In the case of Bianchi space--times, the Einstein equations are
a system of ODE's and the asymptotic behavior of the space--time at the
singularity can be studied using dynamical systems methods. 
It has recently been proved by H. Ringstr\"om \cite{ringstrom:blowup} 
that the singularity in generic 
Bianchi class A space--times with non--stiff matter, 
which includes Bianchi IX, is oscillatory. In the stiff fluid or scalar field 
case, Bianchi
IX is non--oscillatory \cite[\S 20]{ringstrom:attractor}. 
In either case, curvature blows up at the singularity,
which has as a consequence that the space--time is
inextendible, i.e. the conclusion of the cosmic censorship conjecture
cf. \cite{penrose:unsolved}, holds for Bianchi class A.

For generic space--times, in the case of non--oscillatory singularities, 
the BKL proposal leads to the
space--time being asymptotically Kasner at the singularity, with the Kasner
exponents depending on the ``point on the singularity''. In the
oscillatory case, the proposal indicates that locally near the singularity,
the space--time returns infinitely often
to a state which is approximately a non--flat Kasner. From this it would 
follow
that certain curvature scalars such as the Kretschmann scalar $\kappa = \aR_{\alpha
\beta\gamma\delta} \aR^{\alpha \beta\gamma\delta} $ become unbounded near the
singularity. Thus, verifying the correctness of the BKL proposal would go a long way towards
solving the
cosmic censorship problem.

Barrow \cite{barrow78} discussed the structure of the early universe for
space--times with stiff matter and used the term 
``quiescent cosmology'' for this picture. Here we will use the term
``quiescent singularity'' for a singularity with non--oscillatory behavior
induced by the presence of matter. In section \ref{sec:AR} below 
I will report on joint work with
Alan Rendall \cite{andersson:rendall}, where we construct families of four  
dimensional space--times with stiff fluid or scalar field matter, with
quiescent singularities. These families have the same number of free
functions as the full space of solutions to the Einstein equations, and
therefore this result supports the BKL proposal described above for this
case. 

There are important situations where classes of space--times with symmetry
exhibit non--oscillatory behavior at the singularity, even without the
presence of stiff matter. An example is the Gowdy class of space--times
where both numerical and analytical work gives support to the
notion that generic Gowdy space--times have non--oscillatory singularities. 
Gowdy space--times have a 2--dimensional symmetry group $G_2$, with the
action of $G_2$ generated by spatial Killing fields with vanishing twist. 
Rendall and
Kichenassamy used Fuchsian methods to construct families, with the ``right''
number of degrees of freedom, of Gowdy
space--times with non--oscillatory singularity. This result uses a singular
version of the Cauchy-Kowalewski theorem, and thus requires real analyticity for
the data. The Fuchsian method is also used in the four dimensional 
results described in section \ref{sec:AR}. 
The analyticity condition has recently been removed for the Gowdy 
case by Rendall 
\cite{rendall:cinfty}. 
In general,
$G_2$ space--times with matter or with non--vanishing twist 
have oscillatory
singularities \cite{isenberg:etal:inhomog}. 

The dimension of the space--time is significant. It has been argued, using a
BKL type analysis, that in space--time dimension $D \geq 11$, vacuum gravity
generically has non--oscillatory singularity
\cite{demaret:etal:nonoscillatory}.
A low energy limit of string theory leads to consideration 
of Einstein equations in $D$ dimensions, $D=10,11$, 
with matter consisting  of axions,
dilatons (scalar field) and form fields. This model is sometimes called
``stringy gravity''. The possibility that these space--times have 
non--oscillatory singularities
was exploited in the so--called pre--big--bang scenario, where one views the
present universe as arising out of a (future) cosmological singularity 
in a previous universe, the present universe being blown up to a 
scale
determined by the dilaton. This was studied by Buonnanno et al 
\cite{buonnanno:etal:preBB} where it was shown using formal expansions that the
singularities of stringy gravity in spherical symmetry are non--oscillatory. 
Families of stringy gravity solutions with Gowdy symmetry and
non--oscillatory singularity have been constructed using the
Fuchsian method \cite{narita:etal:string}. 
Recent work by Damour and Henneaux
\cite{damour:henneaux:homog,damour:henneaux:cosm} indicates that stringy gravity without
extra symmetry does in fact have oscillatory singularities.

\section{Quiescent singularities} \label{sec:AR}
Here we describe the main results of \cite{andersson:rendall}. For simplicity
we restrict the discussion here to the case of massless scalar field matter.
We consider four--dimensional space--times with metric of the form 
$$
ds^2 = -dt^2+g_{ab}(t)\theta^a \otimes\theta^b , 
$$
so that $t$ is a Gaussian time coordinate and $\{\theta^a\}$ is a coframe dual to
a frame $\{e_a\}$. 
In this case, the second fundamental form of a hyper-surface 
$t=$constant is given
by $k_{ab}=-\half \d_t g_{ab}$.
The Einstein field equations are 
$$
\aR_{\alpha\beta} = 8\pi \nabla_\alpha\phi\nabla_\beta\phi , 
$$
which 
can be written in the following equivalent $3+1$ form. The constraints 
are:
\begin{align*}
R-k_{ab}k^{ab}+(\tr k)^2&=8\pi[ (\d_t\phi)^2+\nabla^a \phi \nabla_a \phi] , 
 \\
\nabla^a k_{ab}-\nabla_b \tr k&=- 8\pi \d_t\phi \nabla_b \phi ,
\end{align*}
and the evolution equations are:
\begin{align*}
\d_t g_{ab}&=-2k_{ab} ,
\\
\d_t k^a{}_b&=R^a{}_b+(\tr k)k^a{}_b-8\pi \nabla^a \phi \nabla_b \phi .
\end{align*}
Here $R$ is the scalar curvature of $g_{ab}$ and $R_{ab}$ its Ricci tensor.
It follows from the Einstein--scalar field 
equations as a consequence of
the Bianchi identity that $\phi$ satisfies the wave equation
$\ame^{\alpha\beta}\nabla_\alpha\nabla_\beta\phi=0$. This has the
$3+1$ form:
$$
-\d_t^2\phi+(\tr k)\d_t\phi+\Delta\phi=0  .
$$
The constraints and evolution equations are together equivalent to the
full Einstein--scalar field equations. 

As mentioned in the introduction, generic non--oscillatory singularities are
expected to have the feature that spatial derivatives become unimportant near
the singularity. We will make an
ansatz for $g_{ab}, k_{ab}, \phi$ which has leading order terms $\og_{ab},
\ok_{ab}, \ophi$ solving a
version of the Einstein evolution equations and the wave equation with
(most) spatial derivatives cancelled. This is the velocity dominated system, with
constraint equations 
\begin{align*}
-\ok_{ab}\ok^{ab}+(\tr \ok)^2&=8 \pi (\d_t\ophi)^2  , 
\\
\nabla^a (\ok_{ab})-e_b(\tr \ok)&=- 8\pi\d_t\ophi e_b(\ophi) , 
\end{align*}
and evolution equations 
\begin{align*}
\d_t \og_{ab}&=-2\ok_{ab} , 
\\
\d_t \ok^a_{\ b}&=(\tr \ok)\ok^a_{\ b}-8\pi \d_t\ophi^2  \delta^a_{\ b}   .
\end{align*}
The velocity dominated scalar field  $\ophi$ satisfies the equation
$$
-\d_t^2(\ophi)+(\tr \ok)\d_t\ophi=0 . 
$$
The velocity dominated evolution equations 
are ordinary differential equations, while the velocity dominated constraint
equations still include 
partial differential equations. The space of solutions of 
the velocity dominated constraint equations can be proved to have 
the same number of free
functions as the standard Einstein constraint equations. 

The solution of the velocity dominated evolution equations is, after an
appropriate choice of time, given by 
$\ok^a_{\ b} = - t^{-1} K^a_{\ b}$, where $K^a_{\ b}(x)$ is a time--independend
symmetric matrix, which is assumed to have positive eigenvalues $\{p_a(x)\}$, 
the
Kasner exponents. The velocity dominated scalar field $\ophi$ is of the form 
$\ophi = A \log(t) + B$, where $A,B$ are time--independent functions.

There is an important technical problem when the matrix $K^a_{\ b}$ has
double eigenvalue at some point $x$. In this case, it is impossible in
general to choose a smooth frame $\{e_a\}$ 
which diagonalizes $K^a_{\ b}$. This causes 
some difficulties which which we will gloss over in the discussion
below. 

Given the solution to the velocity dominated system, we make the following
ansatz for the solution of the full 3+1 Einstein scalar field system. We work 
in terms of a frame $\{e_a\}$ which (approximately) diagonalizes $K^a_{\ b}$, 
let $q_a$ be (approximate) eigenvalues of $K^a_{\ b}$, see 
\cite[\S 5]{andersson:rendall} for details. 

For $s \in \Re$, let $(s)_+ = \max(s,0)$, and let $\alpha_0$ be a small
positive number, to be chosen. Let $\alpha^a_{\ b} = (q_b- q_a)_+ +
\alpha_0$. The ansatz we use is
\begin{align*}
g_{ab}&=\og_{ab}+\og_{ac}t^{\alpha^c_{\ b}}\gamma^c_{\ b} , 
\\
k_{ab}&=g_{ac}(\ok^c_{\ b}+t^{-1+\alpha^c_{\ b}}\kappa^c_{\ b}) , 
\\ 
\phi &= \ophi+t^{\beta}\psi , 
\end{align*}
where $\gamma^c_{\ b}, \kappa^c_{\ b}, \psi$ are $\ordo(1)$. The ordo notation used 
here, and in Theorem \ref{maintheorem} below, 
has a precise technical meaning, cf. \cite[\S 4]{andersson:rendall}.
It is important to
note that $g_{ab}$ is not a priori symmetric but this has to be shown as 
a consequence of the equations. 

Rewriting the Einstein and scalar field evolution equations using the ansatz
gives, among other equations, the following evolution equations for $\gamma^a_{\ b},
\kappa^a_{\ b}$,
\begin{align*}
t\d_t \gamma^a{}_b&+\alpha^a_{\ b} \gamma^a_{\ b}+2\kappa^a_{\ b}
+2\gamma^a_{\ e}(t\ok^e_{\ b})-2(t\ok^a_{\ e})\gamma^e_{\ b}=
-2t^{\alpha^a_{\ e} + \alpha^e_{\ b} 
- \alpha^a_{\ b}} \gamma^a_{\ e}\kappa^e_{\ b} , 
\\ 
t\d_t\kappa^a_{\ b}&+\alpha^a_{\ b}\kappa^a_{\ b}
-(t\ok^a_{\ b})(\tr\kappa)
=t^{\alpha_0}(\tr\kappa)\kappa^a_{\ b} 
+t^{2-\alpha^a_{\ b}} ( \SR^a_{\ b}- 8\pi g^{ac} e_c(\phi) e_b(\phi ) )  . 
\end{align*}
Here $\SR^a_{\ b}$ is the Ricci tensor with respect to a symmetrized version
of $g_{ab}$. It can be shown that the
power of $t$ occurring on the right hand side of 
the evolution equation for $\gamma^a_{\ b}$
is positive. 

In order to write the system in first order form, it is necessary to
introduce an additional field 
$\lambda^a_{\ bc}$, defined in terms of 
the first order spatial derivatives of $\gamma^a_{\
b}$. Proceeding in the
same way for the wave equation results in a first order system in the
variables $\psi, \omega_a ,  \chi$, where $\omega_a, \chi$ encodes first
order space and time derivatives of $\psi$, respectively. 
Let $u =  (\gamma^a_{\ b} , \lambda^a_{\ bc} , \kappa^a_{\ b}, \psi, \omega_a 
, \chi)$ be the unknown in the resulting first order system. 
The Einstein scalar field system then takes the Fuchsian form 
\begin{equation}\label{eq:fuchs}
t \partial_t u + A U = f[t,x,u, u_x] ,
\end{equation}
where 
$f = \ordo(1)$, in the sense that $f$ applied to the fields specified by the
ansatz tends to zero as a power of $t$, as $t$ tends to zero, 
and the matrix function $A$ satisfies a spectral condition,
related to being positive definite, cf. \cite[\S 4]{andersson:rendall}. 
Under an analyticity condition, existence and uniqueness for the system
(\ref{eq:fuchs}) holds by a singular version of the Cauchy--Kowalewski
theorem, cf. \cite[Theorem 3]{andersson:rendall}. The main technical
difficulty in proving that the Einstein--scalar field system is in the
Fuchsian form
(\ref{eq:fuchs}) lies in the proof that the
function $f$ is $\ordo(1)$ in the appropriate sense.  
For the proof that $f$ is $\ordo(1)$, 
the most important point is to estimate the term 
$t^{2-\alpha^a_{\ b}} \SR^a_{\ b}$ occurring in the evolution equation for
$\kappa^a_{\ b}$. This estimate requires a detailed analysis of the blowup
rate of the connection forms of the metric $g_{ab}$, and
their derivatives. The Einstein--scalar field and Einstein--stiff fluid
systems can be written in Fuchsian form only 
if $p_a > 0$, in the
four--dimensional case we are considering. 
The above proves existence and uniqueness of solutions up to the singularity,
for 
the Einstein--scalar field system with real analytic data. 
\begin{theorem}[\protect{\cite[Theorem 1]{andersson:rendall}}]\label{maintheorem}
Let $S$ be a real analytic mani\-fold of dimension 3, and let 
$(\og_{ab}(t),\ok_{ab}(t),\ophi(t))$ be a real analytic 
solution of the velocity dominated Einstein-scalar field equations on $S
\times (0,\infty)$, 
such that $t\tr \ok=-1$ and $-t\ok^a_{\ b}$ has positive eigenvalues.
Then there exists an open 
neighbourhood $U$ of $S\times \{0\}$ in $S\times [0,\infty)$ and a unique 
real analytic solution $(g_{ab}(t),k_{ab}(t),\phi(t))$ of the Einstein-scalar 
field equations on $U\cap (S\times (0,\infty))$ such that for each compact 
subset $K\subset S$ there are positive real numbers 
$\zeta, \beta, \alpha^a_{\ b}$, 
for which the following estimates 
hold uniformly on $K$:
\begin{enumerate}
\item \label{point:scal1st}
$\og^{ac}g_{cb}=\delta^a_{\ b}+o(t^{\alpha^a_{\ b}})$,
\item
$k^a_{\ b}=\ok^a_{\ b}+o(t^{-1+\alpha^a_{\ b}})$,
\item
$\phi=\ophi+o(t^\beta)$, 
\item
$\d_t\phi=\d_t\ophi+o(t^{-1+\beta})$,
\item
$\og^{ac}e_f(g_{cb})=o(t^{\alpha^a_{\ b}-\zeta})$,
\item \label{point:scallast}
$e_a(\phi)=e_a(\ophi)+o(t^{\beta-\zeta})$.
\end{enumerate}
\end{theorem}
The positivity condition on the eigenvalues together with the velocity 
dominated Hamiltonian constraint imply that the function $A$ occurring in
$\ophi$ has the property that $A^2$ is strictly
positive in the velocity dominated solution. Thus vacuum solutions are
ruled out by the hypotheses of this theorem. See \cite[Theorem 2]{andersson:rendall} for 
an analogous result in the stiff fluid case. See also \cite{rendall:blowup}
for a discussion of the case of Einstein equations coupled to a
nonlinear scalar field. It is an interesting problem to generalize the
results discussed here to $C^{\infty}$ data, as was done for the Gowdy
problem in \cite{rendall:cinfty}.


\providecommand{\bysame}{\leavevmode\hbox to3em{\hrulefill}\thinspace}

\end{document}